\documentstyle[12pt]{article}

\begin{document}

\Large
{\bf Relativistic Deuteron Structure Function}  \\

\normalsize
W.Melnitchouk$^a$,
A.W.Schreiber$^b$ and
A.W.Thomas$^c$

{\em $^a$ Institut f\"ur Theoretische Physik,
          Universit\"at Regensburg,
          D-93040 Regensburg, Germany}

{\em $^b$ Paul Scherrer Institut,
          W\"urenlingen und Villigen,
          CH-5232 Villigen PSI, Switzerland}

{\em $^c$ Department of Physics and Mathematical Physics,
          University of Adelaide,
          South Australia, 5005
          Australia}                    \\

{\bf Abstract:}
We calculate the unpolarised deep inelastic structure function
of a relativistic deuteron within a covariant framework.
An exact treatment of nucleon off-shell effects is shown to
give corrections to the widely-used convolution model,
even in impulse approximation.
Neglecting off-shell effects in the extraction of the neutron
structure function from deuterium data introduces errors of order
$1-2\%$.
\vspace*{2cm}

TPR-94-03 \\
PSI-PR-94-15 \\
ADP-93-227/T143 \\
Physics Letters B, in press \\

\newpage

Because of its weak binding the deuteron is thought to be
the ideal target from which to obtain information about the
deep inelastic structure of the neutron.
An accurate determination of neutron structure functions
is vital for testing various parton model sum rules such as
the Gottfried sum rule, which measures the flavour content
of the proton sea, or the Bjorken sum rule in spin-dependent
scattering.
Recent precision experiments on unpolarised \cite{DATAN}
and polarised \cite{DATAS} deuterium targets have enabled both
of these to be tested more rigorously than ever before.
However, the improved accuracy of the experiments has also
meant that a definitive determination of these sum rules
requires serious consideration of the nuclear corrections
that arise when extracting neutron structure functions
from the deuteron data.
Considerable attention has already been focussed on nuclear
shadowing \cite{METHO,SHADD} and meson-exchange \cite{METHO,KAP}
effects, which are typically a few percent below $x \sim 0.1$.
On the other hand, the most obvious nuclear corrections
are those arising from the fact that nucleons in the
deuteron are bound, and thus off-mass-shell ($p^2 \not= M^2$).
Despite this, a fully consistent treatment of off-shell
effects in the deuteron is still outstanding.

The discussion of nucleon off-shell effects is usually based on
some ansatz relating the deuteron structure function
to a convolution involving the free nucleon structure function
at a shifted value of $x$ or $Q^2$ \cite{WEST,BODEK,CONV}.
Instead, we present a systematic treatment of off-shell
effects starting from the most general Dirac structure of
the off-shell nucleon tensor and a covariant $DNN$ vertex function.
The general formalism was developed in Ref.\cite{OFF}.
Here we show how to extract the usual convolution from the
full calculation --- which is exact in impulse
approximation (c.f. Ref.\cite{ST}).
The practical importance of the approximations needed to arrive
at the convolution formula is estimated within a specific model
for the structure of the nucleon and the $DNN$ vertex.

As shown by Melnitchouk et al. \cite{OFF}, in the Bjorken
limit the most general form of the truncated (off-shell)
nucleon structure function is
$\chi(p,q) \equiv I\ \chi_0$
+ $\not\!\!p\ \chi_1$
+ $\not\!\!q\ \chi_2$,
where $\chi_{i}$ are real functions of $p$ and $q$.
To calculate the complete deuteron structure function one
must also describe the off-shell nucleon---deuteron interaction,
which we write in the form:
${\cal A}(P,p) \equiv I {\cal A}_0(P,p)
                    + \gamma^{\alpha} {\cal A}_{1\alpha}(P,p)$.
Taking the trace of ${\cal A}(P,p) \otimes \chi(p,q)$ and
integrating over $p$ gives the spin-averaged
quark distribution per nucleon in the deuteron:
\begin{eqnarray}
q^D(x)
&=& {1\over 2\pi^2} \int dy\ dp^2
 \left( {\cal A}_0\ \chi_0\
     +\ {\cal A}_1 \cdot p\ \chi_1\
     +\ {\cal A}_1 \cdot q\ \chi_2
 \right)                                           \label{full}
\end{eqnarray}
where $x = -q^2 / 2 P \cdot q$ is the Bjorken scaling variable,
and $y = p \cdot q / P \cdot q$ is the fraction of the light-cone
momentum of the deuteron carried by the nucleon.
In impulse approximation the deuteron structure function involves an
on-shell spectator while the off-shell nucleon is struck.
The $DNN$ vertex of Buck and Gross \cite{BG} is therefore ideal
for this purpose.
For a deuteron at rest, it relates ${\cal A}_0$ and
${\cal A}_{1\alpha}$
(which include the off-shell nucleon propagators) to the
relativistic deuteron wavefunctions:
\begin{eqnarray}
{\cal A}_0
&=& 2 M_D \pi^2\ M\
    \left\{ {\cal C}\ -\ 2 {\cal P}\ +\ {\cal D}
    \right\}                                    \nonumber\\
{\cal A}_{1\alpha}
&=& 2 M_D \pi^2\
    \left\{ \left( p_{\alpha} - {p^2-M^2 \over M_D^2} P_{\alpha}
            \right) {\cal C}\
    \right.                                     \label{Ai}\\
& & \left. \hspace*{1.2cm}
         +\ \left( - p_{\alpha}
                   + { M_D^2 + p^2 - M^2 \over 2 M_D^2 } P_{\alpha}
            \right)
            \left( 2 {\cal P} + { M^2 \over {\bf p}^2 } {\cal D}
            \right)
    \right\}                                    \nonumber
\end{eqnarray}
where
\begin{eqnarray}
{\cal C}
&=& u^2(|{\bf p}|)\ +\ w^2(|{\bf p}|)\
 +\ v_t^2(|{\bf p}|)\ +\ v_s^2(|{\bf p}|)               \nonumber\\
{\cal P}
&=& v_t^2(|{\bf p}|)\ +\ v_s^2(|{\bf p}|)               \label{CPD}\\
{\cal D}
&=& {2 |{\bf p}| \over \sqrt 3 M}
    \left[ u(|{\bf p}|)
           \left( v_s(|{\bf p}|) - {\sqrt 2} v_t(|{\bf p}|) \right)
        +\ w(|{\bf p}|)
           \left ( v_t(|{\bf p}|) + {\sqrt 2} v_s(|{\bf p}|) \right)
    \right].                                            \nonumber
\end{eqnarray}
The functions $u$ and $w$ correspond to the usual
non-relativistic $S$- and $D$-state deuteron wavefunctions,
while $v_s$ and $v_t$ represent the
small singlet and triplet $P$-state wavefunctions
which are purely relativistic in origin.
In our calculation we use the wavefunctions from
Ref.\cite{BG} with pseudoscalar $\pi$ exchange,
although similar results are obtained with wavefunctions
derived from a model with pseudovector $\pi N N$ couplings.
(The momentum {\bf p} is calculated as in Eq.(\ref{p2}) below.)

Of particular interest is the connection between Eq.(\ref{full})
and the usual convolution formula \cite{WEST,BODEK,CONV}, in which
the nuclear structure function is a one-dimensional
convolution of the momentum distribution of a bound nucleon
in the deuteron ($\varphi(y)$ in Eq.(\ref{split}) below)
with the on-shell nucleon structure function ($q^N(x/y)$).
In terms of the truncated nucleon functions
the latter is defined as:
\begin{eqnarray}
q^N(x/y)
&=& 4 M\ \chi_0^{on}
 +\ 4 M^2\ \chi_1^{on}
 +\ 4 p \cdot q\ \chi_2^{on}                    \label{on}
\end{eqnarray}
where $\chi_i^{on}$ are obtained by setting $p^2=M^2$
and $p_T=0$ in the fully off-shell $\chi_i$.
In order to derive such a convolution formula one needs to
factorise $q^N(x/y)$ from the integrand in Eq.(\ref{full}).
This in turn requires that either the functions $\chi_i$ are
proportional to each other, or two of them equal to zero.
Alternatively, the $\chi$ and ${\cal A}$ parts of Eq.(\ref{full})
could be factored if the functions ${\cal A}_i$ satisfied
${\cal A}_0/M\
=\ {\cal A}_1 \cdot p/ M^2\
=\ {\cal A}_1 \cdot q\ / p\cdot q$. From
Eqs.(\ref{Ai}) and (\ref{CPD}) it is evident that this
condition is fulfilled in the $p^2 = M^2$ limit by the term
proportional to ${\cal C}$.
By further separating $\chi$ into on- and off-shell parts,
Eq.(\ref{full}) can be recast in the form:
\begin{eqnarray}
q^D(x) &=& \int_x^1 {dy \over y}\ \varphi(y)\ q^N(x/y)\
        +\ \delta^{({\cal A})} q^D(x)\
        +\ \delta^{(\chi)} q^D(x)                       \label{split}
\end{eqnarray}
where the first term is the usual convolution result,
in which $q^N$ is given by Eq.(\ref{on}), and
\begin{eqnarray}
\varphi(y)
&=& { M_D \over 4 }\ y\
    \int_{-\infty}^{p^2_{max}} dp^2\
    {E_p \over p_0}\ {\cal C}(p).                       \label{phi}
\end{eqnarray}
Here the energy $E_p$ is
\begin{eqnarray}
E_p\ =\ \sqrt{M^2 + {\bf p}^2}\
     =\ { M^2 - p^2 + M_D^2 \over 2 M_D }       \label{p2}
\end{eqnarray}
in the deuteron rest frame, and
$p^2_{max} = y M_D^2 - y M^2/(1-y)$ is the maximum kinematic
value of $p^2$.

The two correction terms in Eq.(\ref{split}) can be identified
with the off-shell and $P$-state components of the relativistic
$DNN$ vertex,
\begin{eqnarray}
\delta^{({\cal A})} q^D(x)
&=& { M_D \over 2} \int_x^1 dy\ \int_{-\infty}^{p^2_{max}} dp^2
    \left\{
    \left[ {1\over 2} \left( 1 - {E_p \over p_0} \right) q^N(x/y)
    \right.
    \right.                                             \nonumber\\
& & \hspace*{1cm}
    \left.
 +\ \left( { E_p \over M_D } \chi_1^{on}\
        -\ { P\cdot q \over M_D^2 } \chi_2^{on}
    \right) (p^2 - M^2)\
    \right] {\cal C}                                    \nonumber\\
&+&
    \left[ - 2 M \chi_0^{on}\ +\ 2 {\bf p}^2 \chi_1^{on}\
           +\ (1 - y - {E_p \over M_D}) P \cdot q \> \chi_2^{on}
    \right]\ {\cal P}                                   \label{delA}\\
&+& \left.
    \left[   M \chi_0^{on}\
          +\ M^2 \chi_1^{on}\
          +\ {M^2 \over {\bf p}^2}
             \left( 1 - y - {E_p \over M_D} \right)
             P \cdot q \> \chi_2^{on}
    \right] {\cal D}
    \right\}                                            \nonumber
\end{eqnarray}
and with the off-shell part of the truncated nucleon
structure function,
\begin{eqnarray}
\delta^{(\chi)} q^D(x)
&=& {1 \over 2\pi^2} \int dy\ dp^2
 \left( {\cal A}_0\ \chi_0^{off}\
     +\ {\cal A}_1 \cdot p\ \chi_1^{off}\
     +\ {\cal A}_1 \cdot q\ \chi_2^{off}
 \right)                                        \label{delchi}
\end{eqnarray}
where $\chi_i^{off} \equiv \chi_i - \chi_i^{on}$.

We should note that the convolution term which we have
identified is not unique.
An alternative definition of $\varphi(y)$ in terms of
${\cal C} - {\cal P}$, with an analogous redefinition
of the $\delta^{({\cal A})} q^D$ term, can also produce the required
proportionality of ${\cal A}_0$ and ${\cal A}_{1\alpha}$ needed
to satisfy the
convolution criterion.
Nevertheless Eq.(\ref{phi}) is the most natural choice and one
that has been followed in many phenomenological treatments.
In particular, $\varphi(y)$ includes the infamous relativistic
flux factor, $(E_p/p_0)\ y$ \cite{FLUX},
and therefore satisfies the normalisation condition
\begin{eqnarray}
\int_0^1 dy\ \varphi(y) &=& 1.
\end{eqnarray}
This condition is necessary in order that (for the valence
component) the convolution term alone preserves the baryon
number of the deuteron.
Since both of the correction terms in Eq.(\ref{split}) are
proportional to either $(p^2 - M^2)$ or the $P$-state
wavefunctions, we expect the $\delta^{({\cal A})} q^D$ and
$\delta^{(\chi)} q^D$ terms to be small.
However, to obtain a quantitative estimate of these
corrections requires a model for the off-shell nucleon
functions $\chi_i$.

Our model for the nucleon structure function is motivated
by the work of Refs.\cite{MODELS,CT,GRV}.
In particular, we suppose that at some relatively low
scale ($Q_0 \sim 400$ MeV) the nucleon is well described
in terms of its valence quarks.
Then, as explained in Ref.\cite{OFF}, we can calculate
the $\chi_i$ in terms of a set of phenomenological vertex
functions, $\Phi^{(S)}(p,k)$, describing the process
$N \rightarrow q (qq)$, where the $(qq)$-pair has spin
$S=$ 0 or 1.
On the basis of bag or constituent quark models we expect
these diquark states to have masses
$m_S \simeq$ 700 (900) MeV for $S=$ 0 (1) \cite{CT}.
For simplicity we choose a single function for each
type of vertex, say $I \Phi^{(0)}$ and
$\gamma_5 \gamma_{\alpha} \Phi^{(1)}$, and
parameterise their momentum dependence in the form
\begin{eqnarray}
\Phi^{(S)}(p,k) &=&
N(p^2) { k^2 \over ( k^2 - \Lambda_S^2 )^{n_S} }.       \label{VF}
\end{eqnarray}
With the above values for $m_S$
(note that as the spectator is on-mass-shell,
$m_S^2 = (p-k)^2$)
the cut-offs, $\Lambda_S$, and exponents, $n_S$,
are chosen to fit the on-shell data.
The function $N(p^2)$ parameterises the nucleon off-shell
dependence of the vertex functions.
For the deuteron, because of the strong peak in the
$p_T$ distribution at small transverse momentum
($p_T \sim 25$ MeV), $N(p^2)$ is well approximated
by a constant.

We could equally well choose to parameterise the distributions
at larger $Q_0^2$, however the identification of $m_S$ with diquark
masses would be lost, as would the use of approximate SU(4) symmetry
to relate the spin-flavour distributions to the valence quark
distributions
(namely
$d_{val} = q_{val}^{(1)}$ and
$u_{val} = \left( q_{val}^{(1)} + 3 q_{val}^{(0)} \right)/2$,
where $q_{val}^{(0)}$ and $q_{val}^{(1)}$ are the distributions with
$S$=0 and $S$=1 diquark spectators, each with its first
moment normalised to unity).
Convergence of the $k^2$ integrations in the functions $\chi_i$
imposes constraints on the exponents $n_S$:
$n_0 > 1$, $n_1 > 1.5$.
The correct valence $d/u$ ratio at large $x$ also requires
that $n_1 \approx n_0 + 1$.
The values $n_{0(1)} =$ 1.3 (2.4) are found to reproduce
the large-$x$ data well when evolved to $Q^2 \sim 5$ GeV$^2$.
The other two parameters, the cut-offs $\Lambda_{0(1)}$,
are 750 (450) MeV.
The fact that $\Lambda_0 < \Lambda_1$ can be understood
in a constituent quark picture from the smaller radius
of the scalar two-quark system.

For the sea component of the nucleon we take a similar form for
the vertex functions, but assume that sea quark distributions
are associated with intermediate spectator states having somewhat
larger masses, $\sim 2 m^{val}_{0,1}$.
(In bag model calculations \cite{MODELS} the sea corresponds roughly
to four partons in the intermediate state, compared with two for
the valence quarks.)
To suppress the large-$k_T$ components of the vertex functions
for the sea we choose $n_{0(1)} =$ 4 (5), which, with cut-offs
$\Lambda_{0(1)} =$ 0.7 (0.7) GeV, give the observed rapid fall-off
at large $x$.
In addition, we parameterise the observed flavour asymmetry in
the proton sea by taking
$d_{sea} - u_{sea} = 2 (1-r) q_{sea}$\ \ ($0 < r < 1$),
subject to the constraint
$\int_0^1 dx\ (d_{sea} - u_{sea}) \approx 0.11$ \cite{DATAN},
where $q_{sea} \equiv (u_{sea} + d_{sea})/2$.

Although at the quark model scale, $Q_0$, the low resolution
means that valence quarks dominate, the work of Gl\"uck et al.
\cite{GRV} suggests the phenomenological need of a small
amount of glue.
For the shape of the input gluon distribution we use their
valence-like parameterisation: $x g(x) \sim x^2 (1-x)^4$.
With the above parameters the second moments of $(u+d)_{val}$ and
$q_{sea}$ are $\approx 83\%$ and $3\%$, respectively, which is
sufficient to saturate the momentum sum rule.
The smaller fraction of momentum carried by gluons here compared
with the analysis of Ref.\cite{GRV} reflects the slightly
smaller value of $Q_0^2$ (= (0.39 GeV)$^2$) in our fit,
and the rapid rise of $\langle x \rangle_g$ with $Q^2$.
In Fig.2 we show the resulting proton structure function,
$F_{2p} =\ x \left( 4 u_{val} + d_{val} \right)/9\
        +\ x (6 r + 4) q_{sea}/9$\ \
(with $r \approx 0.91$),
evolved to $Q^2 = 5$ GeV$^2$.
Clearly the data is very well described over
the entire range of $x$.

Without introducing any other parameters we can now calculate
the deuteron structure function,
$F_{2D} =\ 5 x \left( u^D_{val} + d^D_{val} \right)/9\
        +\ 20 x q^D_{sea}/9$.
The result, evolved to $Q^2 = 5$ GeV$^2$, is also shown in Fig.2.
Here $q^D_{sea} \equiv \left( u^D_{sea} + d^D_{sea} \right)/2$,
and the flavour distributions $q^D$ are defined to be those for
a bound proton in the deuteron, with charge symmetry assumed
for the neutron distributions (c.f. Ref.\cite{CSB}).
The agreement with recent data from SLAC, (reanalysed) EMC-NA2
and NMC $F_{2D}$ data \cite{DDATA} is clearly excellent.

In order to exhibit the effect of binding and Fermi motion,
in Fig.3 we show the ratio of $F_{2D}$ to
$F_{2N} \left( = F_{2p} + F_{2n} \right)$ as a function of
the variable $x_N \equiv 2 x$.
It shows the same characteristic features observed in the
nuclear EMC effect for heavy nuclei \cite{NUCLEAREMC}.
The combined nuclear effect in deuterium, due to binding,
Fermi motion and nucleon off-shellness, is predicted
to be about 5\% at $x_N \simeq 0.6 - 0.7$, which happens to be
similar to that found in Ref.\cite{DT}.
(Reliable predictions for the ratio below $x_N \sim 0.2$ would
require inclusion of additional mechanisms beyond the impulse
approximation considered here.)

Having obtained a good fit for the proton and deuteron structure
functions we are now in a position to test the numerical
importance of the off-shell corrections in Eq.(\ref{split}).
In Fig.4 we plot the ratios of (the valence components of)
the individual terms to the total $F_{2D}$.
Both $q^D(x)$ and the convolution component in Eq.(\ref{split})
are normalised to 1 by choosing $N(p^2)$ in Eq.(\ref{VF})
to be a constant, about 0.7\% (2.6\%) larger for the
$S=0$ (1) vertex than the corresponding normalisation
constant for an on-shell nucleon.
As a result the first moments of $\delta^{({\cal A})} q^D$
(negative) and $\delta^{(\chi)} q^D$ (positive) cancel.
The ratio of the convolution component to the total
deviates from unity by about 1--2\% for $x_N < 0.9$.
(In fact, for $x_N > 1$ the off-shell corrections start
to dominate.)
Although the off-shell effects are numerically small,
in any precision analysis of the neutron structure
function all nuclear effects should be included.

In summary, we have calculated the complete relativistic
deuteron structure function within a covariant formalism,
including the effects due to binding, Fermi motion and
nucleon virtuality.
The off-shell effects associated with the nucleon structure
function as well as with the $DNN$ vertex function are found
to give corrections to the usual convolution formula.
Consequently the common deconvolution procedure
\cite{BODEK,FSREP} of extracting the neutron structure function
from deuterium data will introduce errors in $F_{2n}$
of $\approx 1-2\%$ for $x_N < 0.9$.
Off-shell effects also give rise to corrections to the convolution
formula for the polarised deuteron structure function $g_{1D}$
\cite{DATAS}, and the consequences of these for
$g_{1n}$ are currently under investigation
\cite{OFFSPIN}.


We would like to thank Yang Lu for helpful comments
and a careful reading of the manuscript.
This work was supported by the BMFT grant 06 OR 735
and by the Australian Research Council.


\newpage

Figure Captions.

1. Deep inelastic scattering from a deuteron in the impulse
         approximation.
         The respective photon ($q$), quark ($k$), nucleon ($p$)
         and deuteron ($P$) momenta are indicated.

2. Total proton and deuteron $F_2$ structure functions,
         calculated and evolved to $Q^2 = 5$ GeV$^2$.
         The data are a compilation of the SLAC, re-analysed EMC-NA2
         and NMC data \protect\cite{DDATA}
         for $4 < Q^2 < 6$ GeV$^2$.

3. Ratio of the deuteron and isoscalar nucleon
         structure functions in the valence dominated region
         at $Q^2 = 5$ GeV$^2$.

4. Off-shell nucleon and $DNN$ vertex corrections,
         $\delta^{(\chi)} q^D$ (dotted) and
         $\delta^{({\cal A})} q^D$ (dashed), resectively,
         as a ratio to the total $q^D$
         (see Eq.(\protect\ref{split})).
         The solid curve is (convolution/total) $- 1$.
         Note that even though the first moments
         of $\delta^{(\chi)} q^D$ and $\delta^{({\cal A})} q^D$
         are equal in magnitude, because these are divided by the
         $x$-dependent $q^D$, the areas under the
         dashed and dotted curves are not equal.

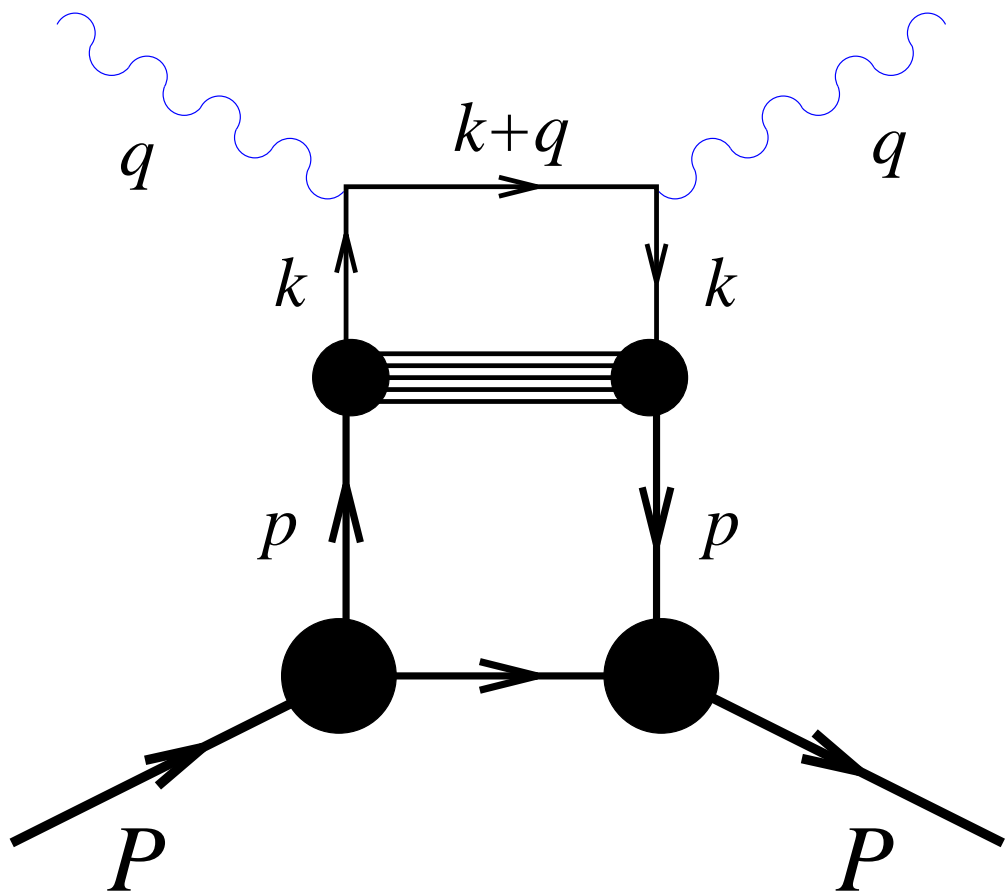
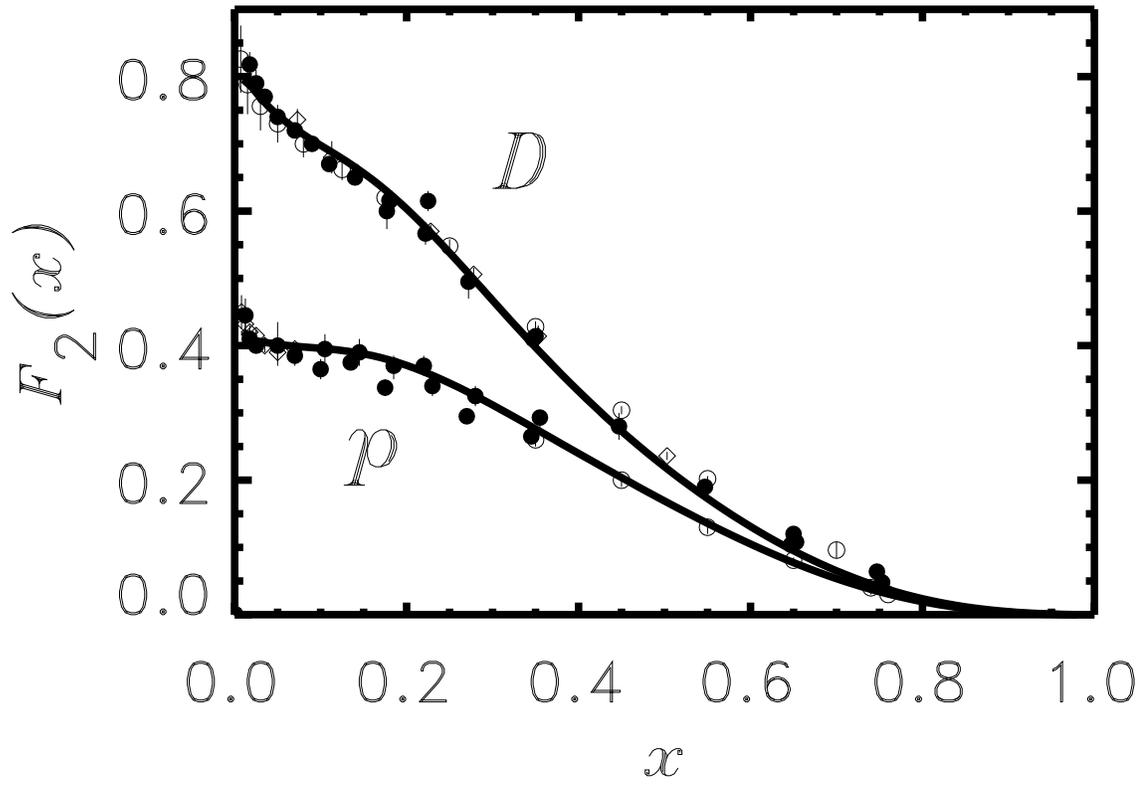
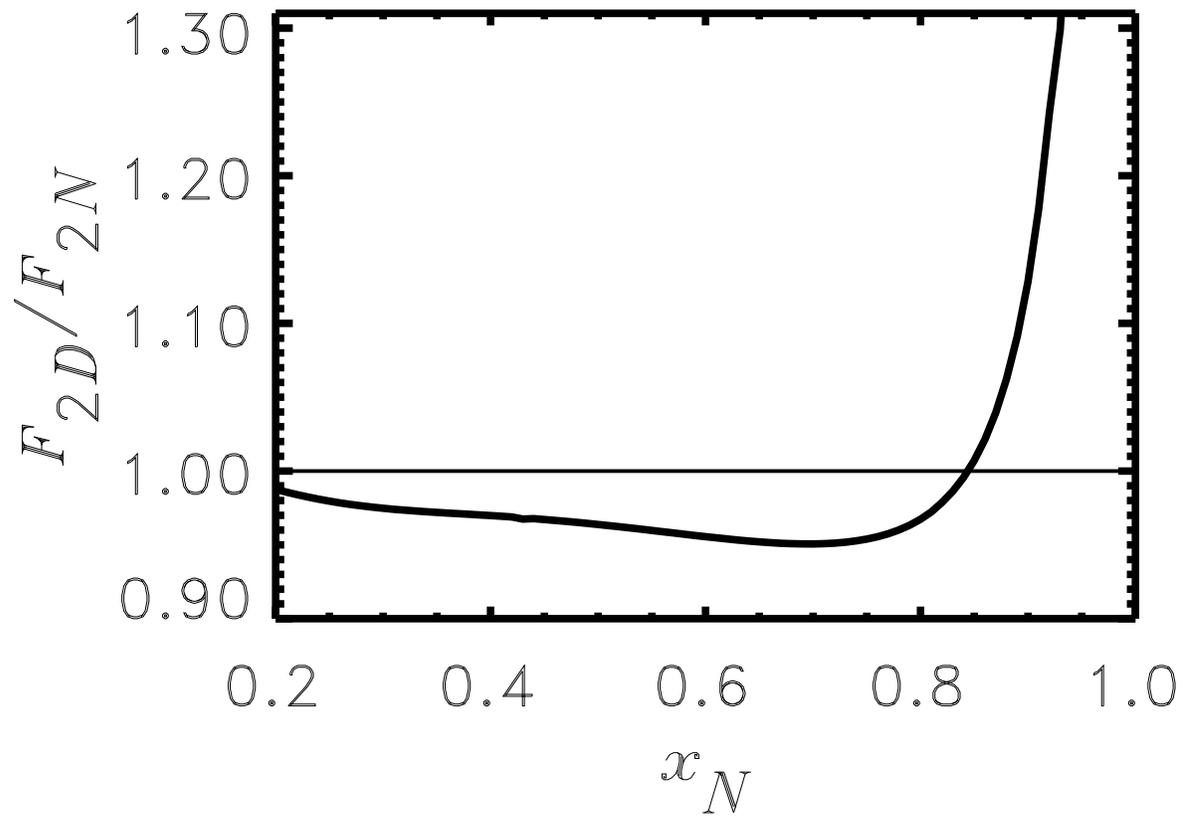
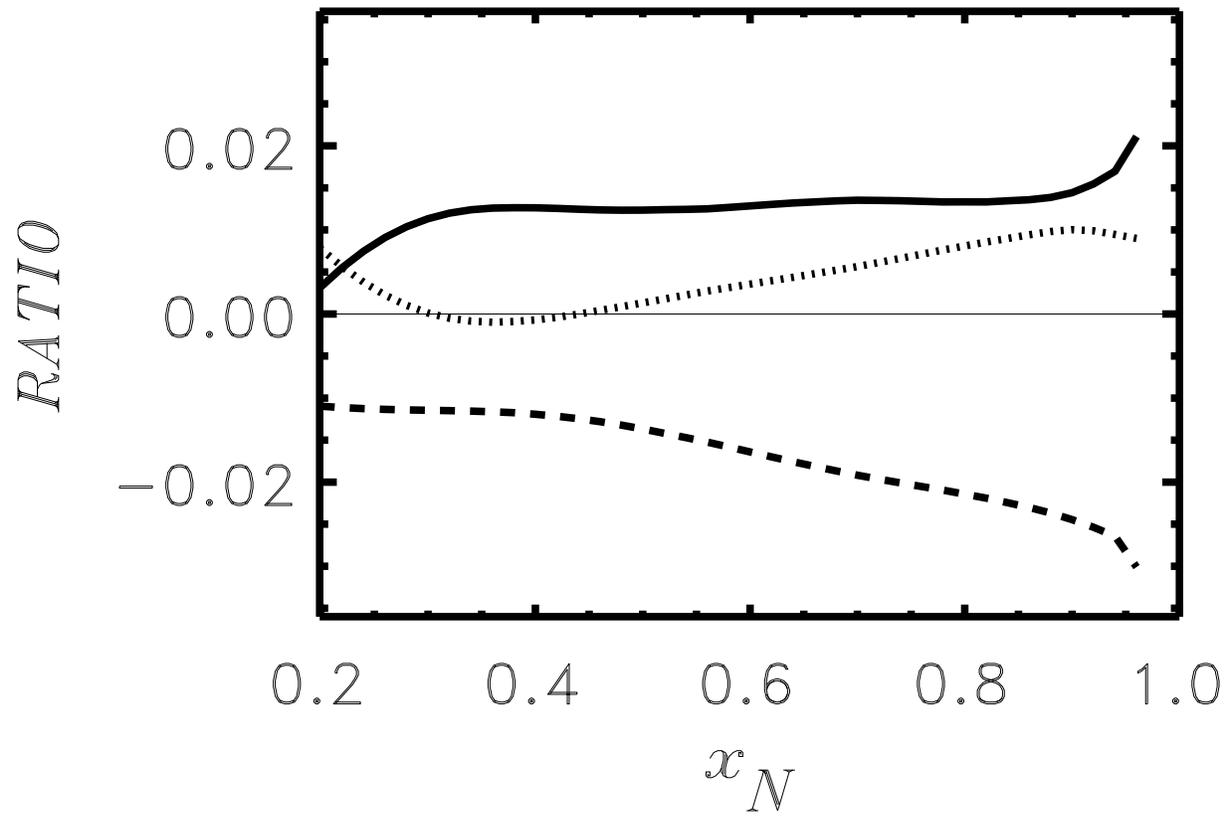
\end{document}